\documentclass{aa}  

\usepackage{graphicx}
\usepackage{txfonts}
\begin{document} 

\title{Deep operator neural network applied to efficient computation of asteroid surface temperature and the Yarkovsky effect}
\author{{Shunjing Zhao\inst{1,2,3}},
          {Hanlun Lei\inst{1,2}},
          {Xian Shi\inst{3}}
          }
\institute{School of Astronomy and Space Science, Nanjing University, Nanjing 210023, China\and
Key Laboratory of Modern Astronomy and Astrophysics in Ministry of Education, Nanjing University, Nanjing 210023, China\and
Shanghai Astronomical Observatory, Chinese Academy of Sciences, Shanghai 200030, China
}

   \date{accepted October, 2024}

 
  \abstract
   {Surface temperature distribution is crucial for thermal property-based studies about irregular asteroids in our Solar System. While direct numerical simulations could model surface temperatures with high fidelity, they often take a significant amount of computational time, especially for problems where temperature distributions are required to be repeatedly calculated. To this end, deep operator neural network (DeepONet) provides a powerful tool due to its high computational efficiency and generalization ability. In this work, we applied DeepONet to the modelling of asteroid surface temperatures. Results show that the trained network is able to predict temperature with an accuracy of $\sim$1\% on average, while the computational cost is five orders of magnitude lower, hence enabling thermal property analysis in a multidimensional parameter space. As a preliminary application, we analyzed the orbital evolution of asteroids through direct N-body simulations embedded with instantaneous Yarkovsky effect inferred by DeepONet-based thermophysical modelling.Taking asteroids (3200) Phaethon and (89433) 2001 WM41 as examples, we show the efficacy and efficiency of our AI-based approach.}

\titlerunning{AI-based thermal}
   \maketitle
%

\section{Introduction}\label{sec1:intro}

Surface temperature distribution is an important property of asteroids, and the base for a wide range of thermal-related studies, such as the Yarkovsky and YORP effects. The prediction of surface temperature distributions requires thermophysical modelling that solves heat conduction equations by incorporating both the shape of the asteroid and various thermal and physical properties. 

Generally, thermal models are classified into simple thermal models that consider the heat conduction of spherical asteroids and more comprehensive thermophysical models that consider a broader range of factors \citep{Delbo2015AsteroidTM}. Simple models include the Standard Thermal Model (STM) \citep{Lebofsky1986}, the Fast Rotating Model (FRM) \citep{Lebofsky1989RadiometryAA}, the Near-Earth Asteroid Thermal Model (NEATM) \citep{Harris1998}, the Night Emission Simulated Thermal Model (NESTM) \citep{Wolters2009}. STM and FRM are rarely used at present. See more details about these models in \citet{Delbo2002} and \citet{Delbo2015AsteroidTM}. More sophisticated thermophysical models have been introduced by \citet{Lagerros1996,Lagerros1996-2,Lagerros1997THERMALPO,Lagerros1998THERMALPO}, \citet{Delbo2004}, \citet{Capek2005} and \citet{Rozitis2011}. 

To obtain temperature distribution over the surface of an irregular asteroid, a polyhedral shape model is commonly used that represents the topography with triangular facets. While some studies consider transverse heat conduction to establish three-dimensional (3D) equations, such as the models used by \citet{Davidsson2014}, \citet{Xu2022} and \citet{Nakano2023}, most of the time it is sufficient to solve one-dimensional (1D) heat conduction equations for each facet \citep{Delbo2015AsteroidTM}. Direct numerical simulation based on finite-difference technique is a high-precision method for solving heat conduction equations, but it requires a high amount of computation cost because of iterative calculations performed on a large number of time-space grids \citep{Delbo2015AsteroidTM}. 

High-fidelity thermophysical models are widely applied in the inversion of asteroids' physical properties that are essential for understanding their formation and evolution. For example, an asteroid’s size is related to its collision history \citep{Bottke2005}, the thermal inertia related to the surface material composition \citep{Delbo2007} and so on. The inversion can be realized by comparing modelled thermal emission to the observations. For example, the thermal inertia of the asteroid (101955) Bennu has been studied in detail by combine the thermophysical models with the mid-infrared observations \citep{Muller2012,Emery2014,Yu2015}. The properties of asteroid (1862) Apollo were constrained by combining the thermophysical model with light-curve, thermal-infrared and radar observations \citep{Rozitis2013}. Moreover, with the determined thermal properties, it is possible to calculate the orbital drift caused by Yarkovsky effect and rotational acceleration from the YORP effect. By fitting them to the observed data, we can further determine axis ratios, effective diameter, geometric albedo, thermal inertia and bulk density \citep{Rozitis2013}, such as the studies about the asteroid (6489) Golevka \citep{Chesley2003}, the asteroid (152563) 1992 BF \citep{Vokrouhlick2008}, Bennu \citep{Chesley2014}, and asteroid (3200) Phaethon \citep{Hanus2018}. In particular, Golevka is the first asteroid that is confirmed with orbital drift caused by Yarkovsky effect \citep{Chesley2003}. Yarkovsky and YORP effects are also critical for the assessment of impact risk \citep{Farnocchia2013,Farnocchia2014}, dynamical spreading of the asteroid families \citep{Vokrouhlicky2010,Hanus2013}, and so on (see more details from \citet{Bottke2006} and \citet{Vokrouhlicky2015}).

Studies about Yarkovsky effect usually utilize analytical thermal models (also called linear models) proposed by \citet{Vokrouhlicky19981,Vokrouhlicky1998,Vokrouhlicky1999} due to high computational burden associated with numerical simulations \citep{Rozitis2012,Xu2022}. Linear models consider spherical asteroids with relatively uniform surface temperature distributions that computes simply and fast, but the tradeoff is precision. However, analytical models may cause additional errors, leading to low-level precision. \citet{Chesley2014} studied the orbital evolution of Bennu from 2000 to 2136 under the influence of Yarkovsky effect calculated under three methods: a transverse acceleration in a simple form, linear model, and nonlinear model based on \citet{Capek2005}. Similarly, \citet{Farnocchia2021} studied the ephemeris and impact hazard assessment of Bennu utilizing three methods, including the linear model, two nonlinear models proposed by \citet{Capek2005} and \citet{Rozitis2011} respectively. \citet{Xu2022} compared the drift caused by the Yarkovsky effect of 34 irregular asteroids from analytical model and numerical simulation, showing that it is overestimated by 16$\%$--36$\%$ with analytical model. For those high-precision problems, numerical simulations for thermophysical models are necessary.

When assessing orbital evolution under Yarkovsky effect, it is common to first generate a lookup table of Yarkovsky force against the orbital elements to avoid solving the thermophysical models at each step of orbit integration. If the orbit is frozen, an interpolation function based on true anomaly angle suffices \citep{Chesley2003,Xu2022}. When the evolution of the orbit is incremental, it is an accepted way to compute the Yarkovsky force from the linearized expansion about a central orbit with semi-major axis $a_0$, eccentricity $e_0$ and further generate a lookup table about $a_0+\delta a$, $e_0+\delta e$, and true anomaly angle \citep{Chesley2014,Farnocchia2021}. However, the linear approximation is no longer applicable when there are significant changes in the orbital elements. Without linear approximation, it is impossible to obtain orbital evolution of asteroids through direct three-dimensional N-body simulations embedded with precise Yarkovsky effect because of high computation cost for producing surface temperature at each step with direct numerical simulation.

To resolve the bottleneck of computational cost in direct numerical simulations, we aim to propose a method that can efficiently produce surface temperature distributions and thus enables the study of long-term orbital evolutions of irregular asteroids under the precise Yarkovsky effect. In recent years, physics informed neural network (PINN) \citep{Raissi2019} have successful applications in varieties of problems because of its efficiency in solving ODEs and PDEs \citep{Cai2020,Martin2022,Martin20221,Zobeiry2021,Mathews2023Solving,He2021,Laghi2023}. The basic model of PINN is a fully-connected neural network, embedding physical equations in the loss function. Equation driven is a prominent feature of PINN. However, it requires to repeatedly train the network when the boundary conditions or parameters of the problem's ODEs or PDEs change. Thus the training cost of PINN may not be less than that of traditional numerical simulations. For the current problem discussed in this work, there are many factors influencing the surface temperature distribution of asteroids. A neural network with greater generalization ability is required to address this problem. In this regard, we adopt deep operator neural network (DeepONet) developed by \citet{LuLu2021}, which can handle this kind of problems with variations of boundary conditions and model parameters. It has been successfully applied in various cases \citep{Garg2022,He2023,Branca2024}.

The following content shows how we (i) explore the feasibility of DeepONet as a unified model for computing the surface temperature distribution of asteroids with varying physical parameters, and (ii) explore the possibilities of DeepONet in solving the problems about the parameter space analysis of Yarkovsky acceleration and the orbital evolution of asteroids under precise Yarkovsky effect.

\section{Thermophysical model and Yarkovsky effect}\label{sec2:model}

In this work, the thermophysical model developed by \citet{Capek2005} is adopted, which considers the self-shadowing effect  between facets while neglecting self-heating. It should be mentioned that the method based on neural network can be applied to more complex models, e.g. the model proposed by \citep{Rozitis2011}. 

\subsection{Thermophysical Model}\label{subsec2:tm}

\subsubsection{Heat conduction equation}\label{subsubsec2:tp}

In thermophysical models, the asteroid is regarded as polyhedron consisting of a large number of triangular facets. When the dimension of facets are sufficiently large, the lateral heat conduction can be neglected. Thus, only the perpendicular heat conduction is taken into account. To obtain the global surface temperature, the 1D heat conduction equation is solved for each facet. See more details in \citet{Lagerros1996}, \citet{Capek2005} and \citet{Rozitis2011}. The 1D heat conduction equation reads
\begin{equation}
\label{eqn:diffusion}
\rho C \frac{\partial}{\partial t} T = \kappa \frac{\partial^2}{\partial z^2} T
\end{equation}
where $\rho$ represents the surface density, $C$ is the specific heat capacity and $\kappa$ is the thermal conductivity. For simplicity, the parameters $(\rho,C,\kappa)$ are assumed to be independent from the time, depth and temperature. 

To make the solution unique, boundary conditions have to be applied. One is the surface boundary condition derived from conservation of energy and the other one is about the constancy of temperature at large depth, given by
\begin{equation}
\label{eqn:condition}
\epsilon \sigma T^4_{z=0}-\kappa \frac{\partial}{\partial z} T_{z=0} = (1-A_B) s \psi F_\odot,\;\frac{\partial}{\partial z} T_{z \rightarrow \infty}=0
\end{equation}
where $\epsilon$ is the emissivity, $\sigma$ is the Stefan–Boltzmann constant and $A_B$ is the Bond albedo. The parameter $s$ indicates whether the facet is illuminated by the Sun. $s=1$ if the facet is illuminated and $s=0$ when unilluminated. $\psi$ represents the cosine function of the solar elevation angle and it can be calculated from cosine law $\psi=\mathbf{a}\cdot\mathbf{b}/|\mathbf{a}||\mathbf{b}|$, where $\mathbf{a}$ and $\mathbf{b}$ respectively represent the direction of the Sun and the outward pointing unitary normal vector of facet. Note that $s$ and $\psi$ depend on both the rotation of asteroid and the topography. In particular, $s=0$ if $\psi \leq 0$ (horizon shadows) or the facet is shadowed by other facets (projected shadows). $F_\odot$ is the incident solar radiation flux determined by $F_\odot=S_\odot/r^2$, where $r$ is the heliocentric distance of asteroid and $S_\odot$ is the solar constant. 

For convenience, Eqs. \ref{eqn:diffusion}-\ref{eqn:condition} need to be normalized as follows \citep{Lagerros1996}:
\begin{equation}
\label{eqn:diffusion_guiyi}
\frac{\partial \overline{T}}{\partial \overline{t}} = \frac{\partial^2 \overline{T}}{\partial \overline{z}^2},
\end{equation}
and
\begin{equation}
\label{eqn:condition_guiyi}
\overline{T}^4_{\overline{z}=0}-\Phi \frac{\partial}{\partial \overline{z}} \overline{T}_{\overline{z}=0} = E(\overline{t}),\;\frac{\partial}{\partial \overline{z}} \overline{T}_{\overline{z} \rightarrow \infty}=0 \\
\end{equation}
with the normalized variables $(\overline{z},\overline{t},\overline{T})$ given by
\begin{equation*}
\overline{z} = \frac{z}{l_s},\;\overline{t} = \omega t,\;\overline{T} = \frac{T}{T_e},
\end{equation*}
where $E(t)=s \psi$ is defined in the range [0,1]. The coefficient $\Phi$ is referred to as the thermal parameter (see Eq. 7 in \cite{Spencer1989}), given by
\begin{equation*}
\Phi=\Gamma \sqrt{\omega}/\epsilon \sigma T_e^3,
\end{equation*}
where $\Gamma=\sqrt{\rho C \kappa}$ is called the surface thermal inertia, $\omega$ is rotational angular velocity of the asteroid, $l_s=\sqrt{\kappa/\rho C \omega}$ is called skin depth and $T_e=\sqrt[4]{(1-A_B)F_\odot/\epsilon \sigma}$ is called characteristic temperature \citep{Lagerros1996,Rozitis2011,Yu2015}. 

\subsubsection{Numerical simulation}\label{subsubsec2:ns}

The first step is to compute the solar radiation flux function about time $E(t)=s \psi$. As mentioned above, $s$ and $\psi$ change periodically as the asteroid rotates, meaning that $E(t)$ is also a periodic function of time. Therefore, it is required to determine $E(t)$ in the first period. The calculation of $\psi$ is based on cosine law.

The time-consuming step is to make shadow tests. There are two types of shadows: horizon shadows and projected shadows. The horizon shadows can be determined from the sign (plus or minus) of $\psi$, while the projected shadows are usually difficult to calculate. It can be regarded as a ray–triangle intersection problem, where the triangle is the facet of interest and the ray can be reasonably approximated as the sunlight passing through the center of gravity of the facet. Note that each facet needs to be tested against the remaining facets and the test process needs to be repeated when the position of the Sun changes. To save computation time, it is usual to store the outcomes and retrieve shadow results from a lookup table \citep{Capek2005,Rozitis2011}. In this work, Möller--Trumbore algorithm is adopted \citep{Tomas1997}. 

To solve the 1D heat conduction equation, we need to divide continuous variables into a set of space-time grids, where the $i$-th grid of depth is denoted by $i\Delta \overline{z}$ and the $j$-th grid of time is denoted by $j\Delta \overline{t}$. As a result, Eqs. \ref{eqn:diffusion_guiyi}-\ref{eqn:condition_guiyi} are discretized into the following form \citep{Rozitis2011}:
\begin{equation}
\label{eqn:diffusion_difference}
\overline{T}_{i,j+1}=(1-2\frac{\Delta \overline{t}}{\Delta \overline{z}^2})\overline{T}_{i,j}+\frac{\Delta \overline{t}}{\Delta \overline{z}^2}(\overline{T}_{i+1,j}+\overline{T}_{i-1,j}),
\end{equation}
and
\begin{equation}
\label{eqn:condition_difference}
\overline{T}_{0,j}^4-\Phi \frac{\overline{T}_{1,j}-\overline{T}_{0,j}}{\Delta \overline{z}}=E(j \Delta \overline{t}),\;\overline{T}_{n,j}=\overline{T}_{n-1,j}.
\end{equation}
Note that the surface boundary condition can be solved to determine $\overline{T}_{0,j}$ from the known $\overline{T}_{1,j}$ by taking advantage of Newton--Raphson iterative algorithm. 

The surface of an asteroid could form a static temperature distribution in an inertial frame, meaning that the temperature is invariant at the subsolar point after a full rotation. This provides a cycle termination condition to solve Eqs. \ref{eqn:diffusion_difference}-\ref{eqn:condition_difference}.

The numerical simulation of determining global surface temperature of an asteroid is a $N^3$ problem, whose scale is determined by the number of facets, steps of iterations and the numbers of the time grid. It needs to be repeatedly solved once any model parameter changes, thus taking a large amount of computational resource.

\subsection{Yarkovsky effect}\label{Sunsec2:ye}

Yarkovsky effect includes the diurnal and seasonal components \citep{Bottke2006}. The diurnal Yarkovsky effect maximizes when the spin axis of the asteroid is perpendicular to its orbital plane and the seasonal effect maximizes when the spin axis of the asteroid is in its orbit plane. 

For the diurnal effect, the hotter spot on the surface of an asteroid is shifted away from the subsolar point into the afternoon because of thermal inertia. These regions will reradiate more energy into the space, meaning that more momentum departs from the hotter part of the asteroid as that the infrared photon carries away momentum equal to $E/c$, where $E$ is the energy and $c$ is the speed of light. This mechanism causes the Yarkovsky force along the transverse and radial directions. As for the seasonal effect, its timescale is equal to the orbital period of asteroid. The hemisphere facing the Sun is hotter during the revolution, causing along-track Yarkovsky force as a resistance to the movement in a small eccentricity orbit. This effect is only important for those asteroids with relatively high thermal inertia \citep{Rozitis2012}.

The first step to determine Yarkovsky force is to compute the surface temperature of the asteroid by the numerical simulation mentioned above. Then, the Yarkovsky force is determined by assuming that the emitted radiation follows Lambert’s law, given by \citep{Spitale2001,Bottke2006}
\begin{equation}
\label{eqn:Yarkovsky_force}
\mathbf{F}=-\frac{2}{3c} \epsilon \sigma \int T^4 \mathbf{n} \mathrm{d} S,
\end{equation}
where $\mathbf{n}$ is the external unitary normal vector of the facet and $S$ is the area of the facet. In practical simulations, the integral can be approximated as the sum of the force on each facet.

\section{Deep operator neural network (DeepONet)}\label{sec3:deep}

Numerical simulation discussed in Sect. \ref{sec2:model} is an accurate approach for producing surface temperature distribution of an asteroid. However, its inefficiency will become an obstacle for many projects, especially in those situations requiring repeated computation of surface temperature. In this regard, neural network provides a powerful tool with less computation cost and greater ability to generalize. In this study, the deep operator neural network (DeepONet), developed by \citet{LuLu2021}, is adapted to address the problem. 

\subsection{Architecture of DeepONet}\label{subsec3:nns}

DeepONet is a deep neural network to learn operators, based on the universal approximation theorem of operator \citep{Chen1995,LuLu2021}. The operator about a function of $u$ in space $C(K_1)$ is denoted by $G$, then the corresponding function in another function space $C(K_2)$ can be regarded as $G(u)$. This mapping relation $u \rightarrow G(u)$ can be approximated by the networks with inputting $u$ and outputting $G(u)$. Take $y$ as any variable in the domain of function $G(u)$. $G(u)(y)$ is a real number, and the mapping relation $y \rightarrow G(u)(y)$ can also be approximated by networks. These two parts constitute DeepONet: branch networks for $u \rightarrow G(u)$ and trunk networks for $y \rightarrow G(u)(y)$. In fact, such an approximation can be regarded as the process of learning potential nonlinear operators from physical equations (see more details in \citet{LuLu2021}).

In the present work, the solar radiation flux function $E$ is the key function and the thermal parameter $\Phi$ is the key parameter of boundary conditions for heat conduction equation. When the two factors are determined, the surface temperature distribution is determined by solving the 1D heat conduction equation. As mentioned in previous section, the flux function $E$ varies with the position of the Sun, thus the external unitary normal vector of the facet or the shadow of the facet may change. In addition, the thermal parameter $\Phi$ is related to thermal parameters, including the surface thermal inertia $\Gamma$ (it is a function of the surface density $\rho$, the specific heat capacity $C$ and the thermal conductivity $\kappa$), rotational angular velocity $\omega$, Bond albedo $A_B$, the heliocentric distance $r$ and so on. 

We denote the solution of heat conduction equation by
\begin{equation*}
G(E,\Phi)(\overline{t}, \overline{z}),
\end{equation*}
where $G(E,\Phi)$ is the operator to be learned. The operator is different from the original version of DeepONet. In practice, $E$ is suggested to be represented discretely. A convenient way is to employ the function values evaluated at sufficiently large but finite number of locations $[E(\overline{t}_1),E(\overline{t}_2),...,E(\overline{t}_m)]$, and the variables $[\overline{t}_1,\overline{t}_2,...,\overline{t}_m]$ are called sensors. $\Phi$ is a real number that can be directly input into the network. Therefore, the input of the branch networks are the functions evaluated at discrete locations $[E(\overline{t}_1),E(\overline{t}_2),...,E(\overline{t}_m)]$ and the real number $\Phi$.

The solution of heat conduction equation is a 2-dimension matrix about $\overline{z}$ and $\overline{t}$. Note that we only concern the surface temperature at a certain time for a facet, meaning the temperature at $\overline{z}=0$ and $\overline{t}=0$ is required, so it is appropriate to reduce the range of variable for better approximation and more convenient training. In practice, grid of depth $[\overline{z}_1,\overline{z}_2,...,\overline{z}_n]$ is chosen to be inputs of the trunk networks.

In summary, we can see that branch networks are used to learn the operator associated with parameters of dynamical model and the trunk network is used to learn the mapping relation of input and output associated with the heat conduction equation. Based on this concept, a variety of similar problems can be addressed by taking advantage of DeepONet.

There are $n$ variables in the input of trunk network, meaning that the output also includes $n$ components. Thus, it is necessary to match the numbers of output with branch networks. Two approaches can be employed: one is to repeat the results of the branch network over $n$ times and the other one is to repeat the input locations $[E(\overline{t}_1),E(\overline{t}_2),...,E(\overline{t}_m)]$ and parameter $\Phi$ over $n$ times. The two approaches are equivalent. In this work, the second is taken.  The input after repetition reads
\[
\begin{pmatrix}
E^{(1)}(\overline{t}_1)&E^{(1)}(\overline{t}_2)&\cdots&E^{(1)}(\overline{t}_m) \\
E^{(2)}(\overline{t}_1)&E^{(2)}(\overline{t}_2)&\cdots&E^{(2)}(\overline{t}_m) \\
\vdots&\vdots&\ddots&\vdots \\
E^{(n)}(\overline{t}_1)&E^{(n)}(\overline{t}_2)&\cdots&E^{(n)}(\overline{t}_m) \\
\end{pmatrix}
,\;
\begin{pmatrix}
\Phi^{(1)} \\
\Phi^{(2)} \\
\vdots \\
\Phi^{(n)} \\
\end{pmatrix}.
\]

\begin{figure*}
\centering
\includegraphics[width=0.7\textwidth]{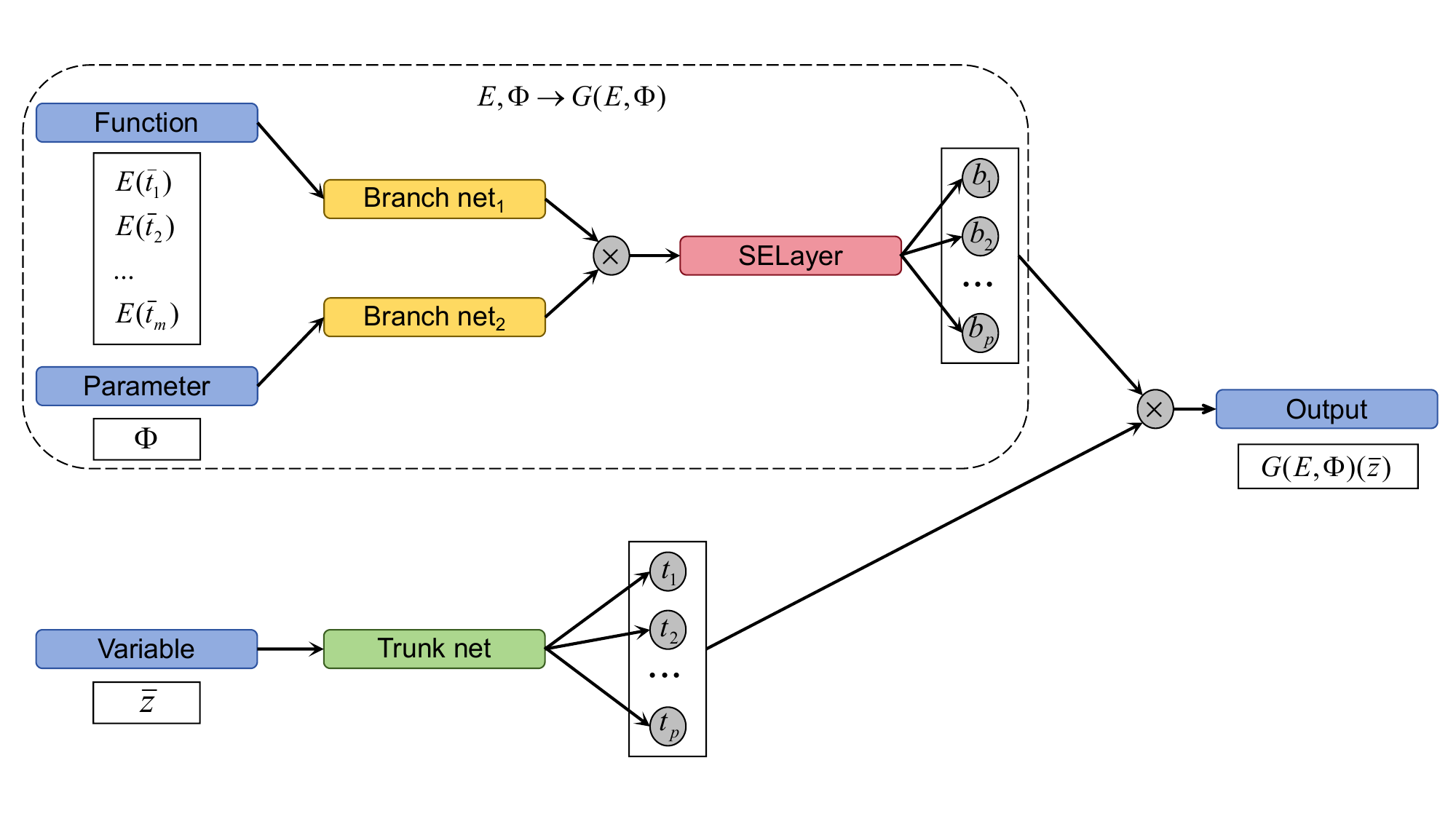}
\caption{Architecture of the modified DeepONet adopted in this work. The network learning the operator $G:(E,\Phi)\rightarrow G(E,\Phi)$ includes three inputs, the function $E$, the parameter $\Phi$. $E$ is presented as some discrete locations with $m$ elements sampled on $[\overline{t}_1,\overline{t}_2,...,\overline{t}_m]$. In this work, $\overline{z}$ limited in a small range is considered as the only variable input to trunk net. The input of SELayer receives the Hadamard product of the outputs from branch net$_1$ and branch net$_2$. The final output is the dot product of $[b_1,b_2,...,b_p]$ and $[t_1,t_2,...,t_p]$ with a bias.}
\label{fig:architecture}
\end{figure*}

\begin{figure*}
\centering
\includegraphics[width=0.7\textwidth]{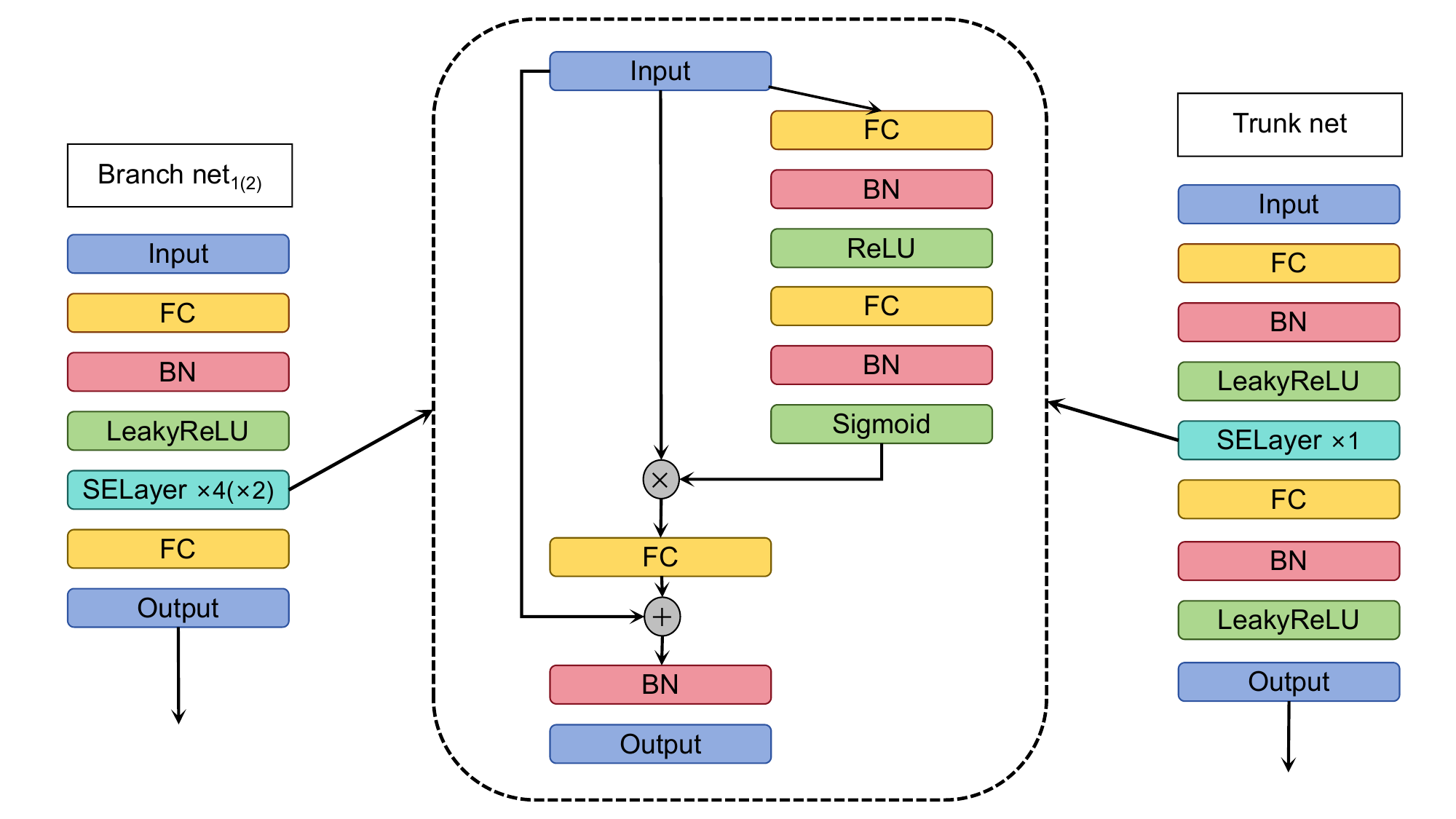}
\caption{Detail architectures of branch net$_1$, branch net$_2$ and SELayer adopted in this work, where FC is the fully-connected layer, BN is the batch normalization layer. The input of branch net$_{1(2)}$ experiences the initial feature extraction through a fully-connected layer and then is carried out the feature amplification by 4(2) SELayer. Trunk net is similar but it should be noted that there is an activation function before output. SELayer in this work is a modified Squeeze-and-Excitation Layer specially for fully-connected architecture.}
\label{fig:detail}
\end{figure*}

The schematic architecture of DeepONet adapted for the problem considered in this work is shown in Fig. \ref{fig:architecture}, where the trunk and branch networks are presented in Fig. \ref{fig:detail}. As one of the improvements to the original DeepONet, two branch networks are used to separately learn the operators related to the radiation flux function $E$ and the thermal parameter $\Phi$. $E$ is input into branch net$_1$ and $\Phi$ is input into branch net$_2$. The Hadamard product between the output of them become the input of SELayer. As a second improvement to the original DeepONet, we introduce the attention mechanism in the SELayer, further improving accuracy of approximation \citep{Vaswani2017,LuLu2021}. The original SELayer is called Squeeze-and-Excitation Networks \citep{Hu2018}. The output is the product between the importance weights and the corresponding channels of the previous feature map. Its essence is to learn the influence of each channel on the final output, thus achieving better feature extraction \citep{Hu2018,Woo2018,Wang2020}. In this work, SELayer is modified by introducing fully-connected architecture, and another fully-connected layer is appended for residual operation following the step of "Excitation". The modified SELayer is applied to branch net$_1$, branch net$_2$ and trunk net.

The output from SELayer is denoted as $b_k$ and the output from trunk net is $t_k$ with $k$ changing from 1 to $p$. The final output is the dot product between them with a bias $b_0$,
\begin{equation}
G(E,\Phi)(\overline{z})=\sum_{k=1}^p b_k t_k+b_0
\label{eqn:output}
\end{equation}
Note that the trunk network applies the activation functions LeakyReLU \citep{Maas2013RectifierNI} in the last layer, thus we can regard the adapted DeepONet network consisting of two branch networks and a trunk network as a full trunk network, where the weights in the last layer are parameterized by branch networks.

\subsection{Dataset and training}\label{subsec3:d}
DeepONet is a data-driven neural network, thus a high-quality training dataset is crucial for its performance. The dataset can be produced with random pairs of $E$ and $\Phi$ through direct numerical simulations. \citet{LuLu2021} has provided two function spaces: Gaussian random field (GRF) and Chebyshev polynomials. In this study, we use mean-zero GRF denoted as
\begin{equation*}
\begin{aligned}
&u \sim \mathcal{G}(0,k_l), \\
&k_l(x_1,x_2)=\exp{(\frac{-||x_1-x_2||^2}{2l^2})},
\end{aligned}
\end{equation*}
where the covariance kernel $k_l(x_1,x_2)$ is the radial-basis function (RBF), called Gaussian kernel in GRF. $l$ is the length-scale determining the smoothness of the function (a larger $l$ corresponds to a smoother $u$), $(x_1,x_2)$ are feature vectors, and $||x_1-x_2||$ represents the Euclidean distance between these two vectors. 

Note that the random functions generated from GRF theoretically extend over large ranges, but $E$ is controlled in the range [0,1], as mentioned in Sect. \ref{subsubsec2:tp}. Therefore, we rescale the random functions by applying the sigmoid function $1/(1+e^{-x})$. In order to ensure the completeness of the function space, we perform translation operations on partial random functions before inputting them into the sigmoid function, meaning that the actual transformation formula is denoted as $1/(1+e^{-(x+d)})$, where $d$ represents the distance of translation along the y-axis. This step helps to maintain endmost values (such as in the range of (0.9,1), (0,0.1), ...). Then the new random functions can be sampled on sensors $[\overline{t}_1,\overline{t}_2,...,\overline{t}_m]$ to obtain the discrete input solar radiation flux functions. 

The dataset used in this work consists of 16000 functions without translation, 8000 functions with $d=-2.5$, 8000 functions with $d=-3$, 6000 functions with $d=-6$, 2000 functions $d=-10$, and 6000 functions $d=6$. There is a corresponding random $\Phi$ for each function. By feeding them into the heat conduction equation and performing numerical simulation we obtain the corresponding temperature distributions that form the overall dataset. Note that the temperature computed from these functions fall within different ranges between 0 and 1, with some having very small differences between the maximum and minimum values. The network may not be very sensitive when learning from these data. So it is important to preprocess the data before training. Z-score normalization is an effective method, reading
\begin{equation*}
Z=\frac{X-\mu_X}{\sigma_X}
\end{equation*}
where $X$ is the original data, $Z$ is the data after transformation, $\mu_X$ is the mean of $X$, and $\sigma_X$ is the standard deviation of $X$. The primary purpose of Z-score is to standardize data of different magnitudes into a single scale and the data measured uniformly by Z-score can ensure comparability among them. 

The loss function in this work is the mean square error (MSE) between the temperature from the dataset and the output from DeepONet, and the optimization algorithm is adaptive moment estimation method (Adam). To avoid the issue of the loss function oscillating widely and converging slowly, we employ the learning rate decay approach. The learning rate in training are diminishing with increasing epochs, reading 
\begin{equation*}
l_r = \frac{l_{r0}}{(1+15\frac{e_p}{e_{pmax}})^{0.75}}
\end{equation*}
where $l_{r0}$ is the initial learning rate, $e_p$ is the current epoch and $e_{pmax}$ is the total epoch. Table \ref{tab:setting} summarizes the settings for the training process. With the dataset prepared, the training time is about 10 minutes.

\begin{table}[]
\footnotesize
\centering
\caption{Parameter settings of training.}
\setlength{\tabcolsep}{4 mm}{
\begin{tabular*}{0.5\textwidth}{@{\extracolsep{\fill}}lc@{\extracolsep{\fill}}}
    \hline
        {Batch size} & 600 \\ \hline
        {Epochs} & 30,000 \\ \hline
        {Initialization} & normal distribution \\ \hline
        {Loss function} & MSE \\ \hline
        {Method} & Adam $(\beta_1=0.8,\beta_2=0.99)$ \\ \hline
        {Initial learning rate} & 0.001 \\ \hline
\end{tabular*}}
\label{tab:setting}
\end{table}

\section{Results}\label{sec4:res}

We use two indicators to assess the performance of the trained DeepONet: the cost of computation time and the level of accuracy.

\subsection{Cost of computation time}\label{subsec4:tce}

Short inference time is one of the most prominent advantages of neural network over numerical simulation. Additionally, the strong generalization ability of neural networks supports practical applications to a variety of problems after training. Under a given hardware, inference time of neural networks depends on the batch size of the input and the intrinsic complexity of the problem.

\begin{figure*}
\centering
\includegraphics[width=0.8\textwidth]{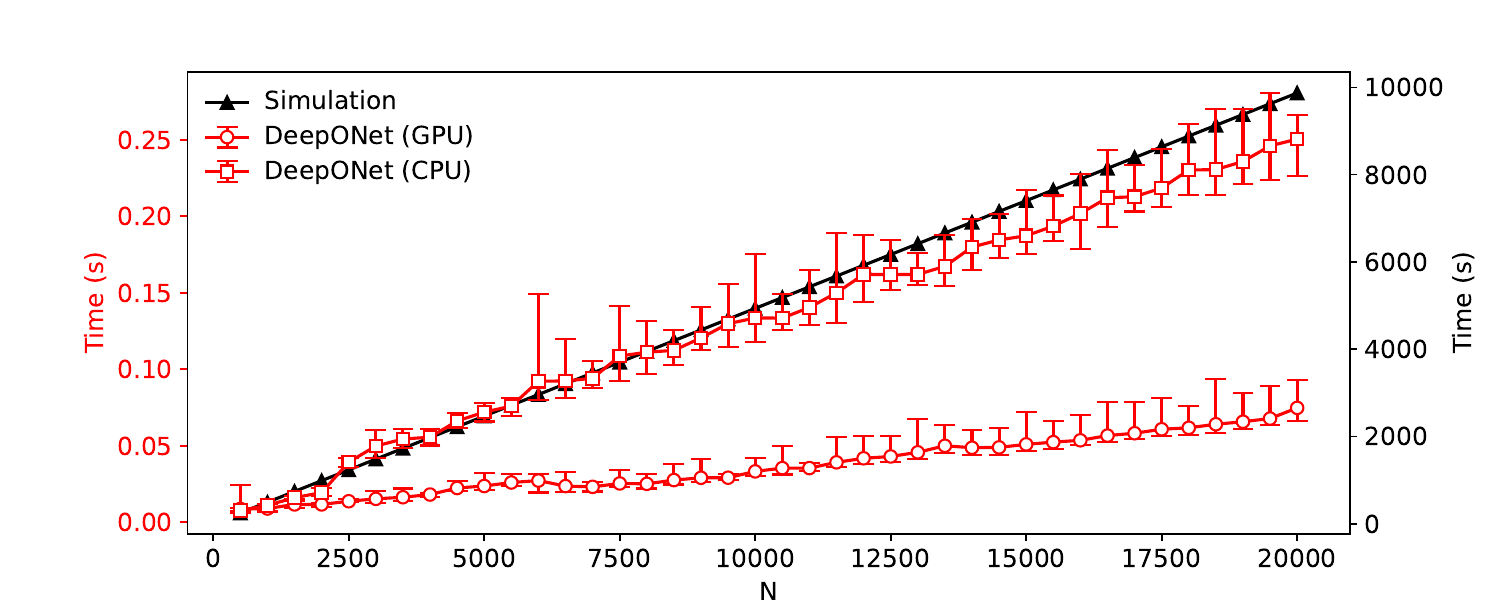}
\caption{Time cost of DeepONet and traditional numerical simulation as a function of number of facets. The red line consisting of error bars centered around circles represents the time cost required by DeepONet in GPU and the red line consisting of error bars centered around squares represents the time cost required by DeepONet in CPU. The black line is the time cost required by traditional numerical simulation.}
\label{fig:timecost}
\end{figure*}

In order to make a direct comparison of the time cost, both the DeepONet algorithm and direct numerical approach (difference algorithm) are adopted to produce the surface temperature for a test sphere-shaped asteroid with the number of facets varying in the range of [500,20000]. Fig. \ref{fig:timecost} shows the time consumption of three methods as a function of the number of facets. For each set of temperature calculation, we perform DeepONet's computation over 20 times using both GPU and CPU. The average computation time with GPU and CPU is shown by hollow circles and hollow squares respectively, and the maximum and minimum time are marked in the form of error bar. Regarding numerical simulations, we pre-calculate the temperatures for a large number of different facets and obtain the average time cost per facet, the results shown in Fig. \ref{fig:timecost} represent such an average value multiplied by the number of computations. All computations are performed in a personal computer\footnote{laptop equipped with AMD Ryzen 9 6900HX with Radeon Graphics and NVIDIA GeForce RTX 3070Ti Laptop GPU} and under the environment of Python 3.8. 

As shown in Fig. \ref{fig:timecost}, DeepONet plus GPU proves to be the most efficient strategy. For an asteroid with 20,000 facets, DeepONet plus GPU can produce the global temperature distribution in less than 0.06 seconds, over 160000 times faster than the numerical simulation (at least five orders of magnitude improvement in terms of computational efficiency). DeepONet plus CPU is also significantly faster than numerical methods, but shows less advantage than the GPU solution with a large number of facets. Furthermore, if we need to compute a large number of global temperature distributions, such as for different asteroids or different parameters, vectorized operations can be further used to ensure that the computation time does not increase linearly with the number of the global temperature distributions, because the internal calculations of networks are carried out using matrix operations. 

\subsection{Performance of accuracy}\label{subsec4:ea}

We take the example of three asteroids with shapes of sphere, biaxial ellipsoid ($a_x=2a_y=2a_z$) and triaxial ellipsoid ($a_x=2a_y=3a_z$), which can represent widely used convex polyhedral shape models of asteroids.

\begin{figure*}
\centering
\includegraphics[width=0.8\linewidth]{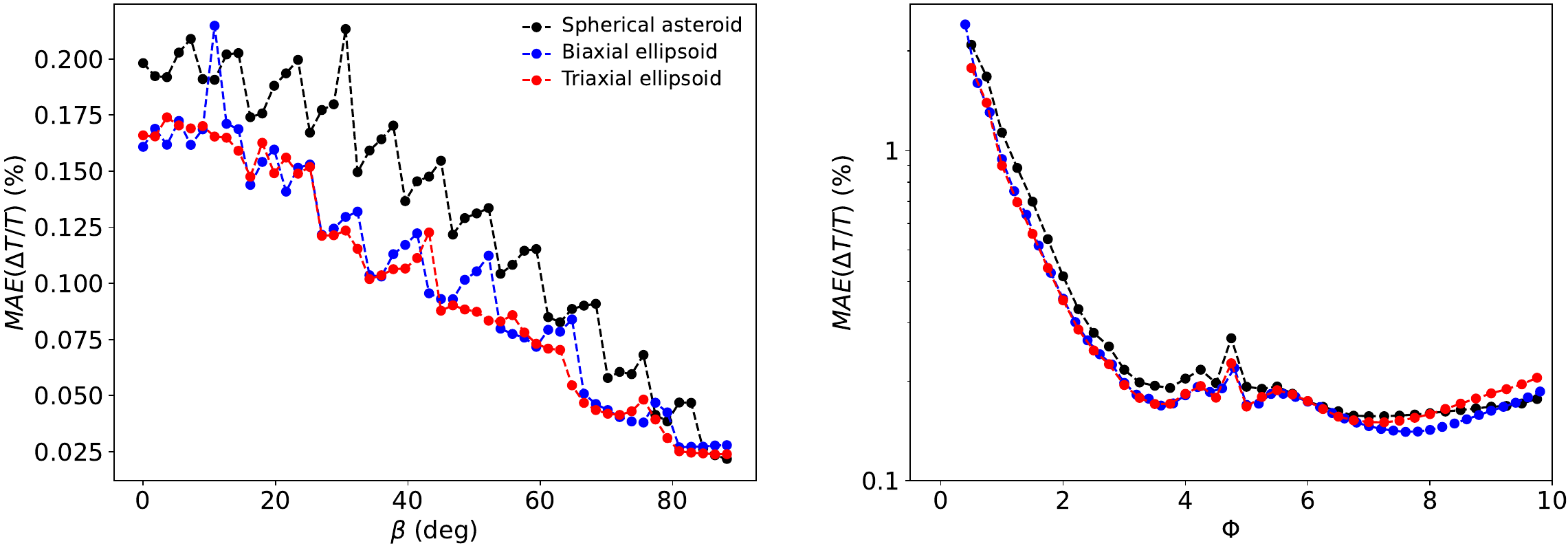}
\caption{Error between the results from DeepONet and numerical simulation. The panels describe the mean absolute error of $\Delta T/T$ as a function of $\beta$ (left panel) and of $\Phi$ (right panel), where the black lines represent the spherical asteroid, the blue lines are the biaxial ellipsoid and the red lines are the triaxial ellipsoid. The left panels fix $\Phi=5$ and the right panels fix $\beta=90^{\circ}$.} 
\label{fig:error}
\end{figure*}

We calculate the global temperature distribution for different obliquity $\beta$, defined as the angle between the spin axis and the normal vector of the orbital plane, and different thermal parameter $\Phi$. Fig. \ref{fig:error} shows the result. The mean absolute error (MAE) stands for the relative deviation of temperature $\Delta T/T$ between DeepONet and numerical simulation. 

The overall error is maintained at a relatively low level of less than 2\%. We noticed certain patterns in the error distribution: (a) The error decreases, in general, as $\beta$ increases. It indicates more accurate temperature in the high-temperature areas, since the number of high-temperature facets increases and the highest temperature is also higher when $\beta$ is larger. (b) The error decreases as the thermal parameter increases, which similarly implies more accurate temperature in the high-temperature areas, since the overall temperature distribution becomes more uniform, the highest temperature decreases but the number of high-temperature facets increases when $\Phi$ is larger (also meaning that the surface thermal inertia or the rotational angular velocity is larger). The conclusion is helpful for calculating the Yarkovsky force, because the influence on the force is greater in high-temperature regions ($F$ is directly related to $T^4$).


\begin{figure*}
\centering
\includegraphics[width=0.84\linewidth]{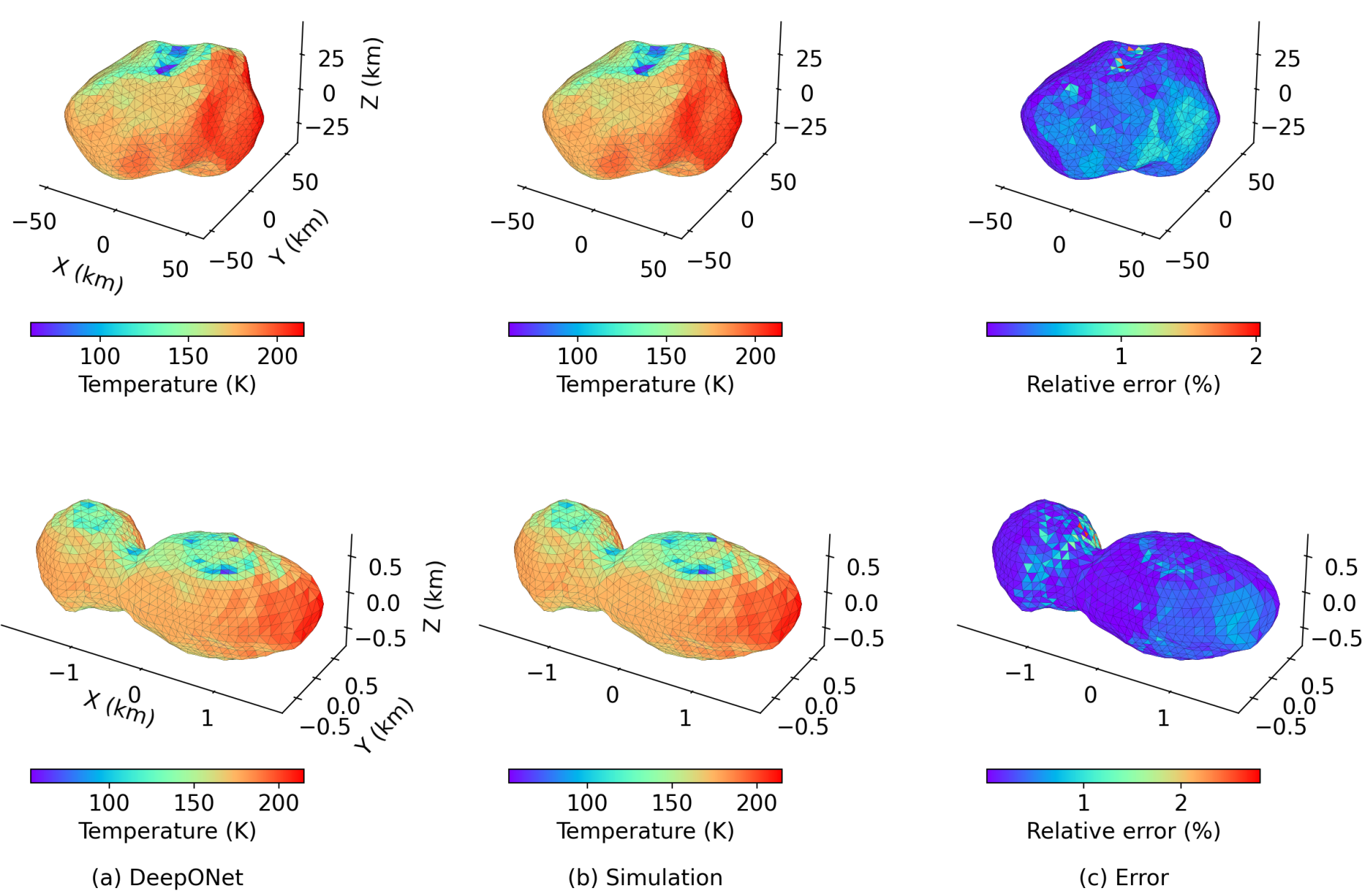}
\caption{Surface temperatures of two irregular asteroids from DeepONet and traditional simulation, and the relative errors between them. The top-row panels are for a main belt asteroid (Lutetia) and the bottom-row panels are for a Amor-class asteroid (1996 HW1). The left-column panels describe the results from DeepONet, the middle-column panels are the results from traditional simulation and the right-column panels are relative errors between the two methods.}
\label{fig:irregularity}
\end{figure*}

Irregular asteroids require consideration of projected shadows between arbitrary two facets, resulting in a more complex radiation flux function. It is a great challenge for the generalization ability of the network. To analyze the accuracy performance of DeepONet on irregular asteroids, we take as examples the dataset of asteroids (21) Lutetia and (8567) 1996 HW1 from the Planetary Data System\footnote{https://sbn.psi.edu/pds/shape-models/}. The thermal parameters are assumed as $\Phi=6.4163$ and $T_e=258.3404\, \mathrm{K}$, calculated by some assumed physical parameters ($A_B=0.33$, $\epsilon=0.9$, $\Gamma=400$, $r=2$, $P=8.76243$, the units are the same as Table \ref{tab:paraspace} and \ref{tab:dpa}). These two irregular asteroids contain various degrees of projected shadows: some are the large-scale shadows caused by the concave shape, and some are small-scale shadows due to the local topography, which represent the real situation of most asteroids. Fig. \ref{fig:irregularity} shows the comparison between the global temperature distributions of the two asteroids from DeepONet and numerical simulation, with the Sun located in the positive x-axis direction.

\begin{figure*}
\centering
\includegraphics[width=0.8\linewidth]{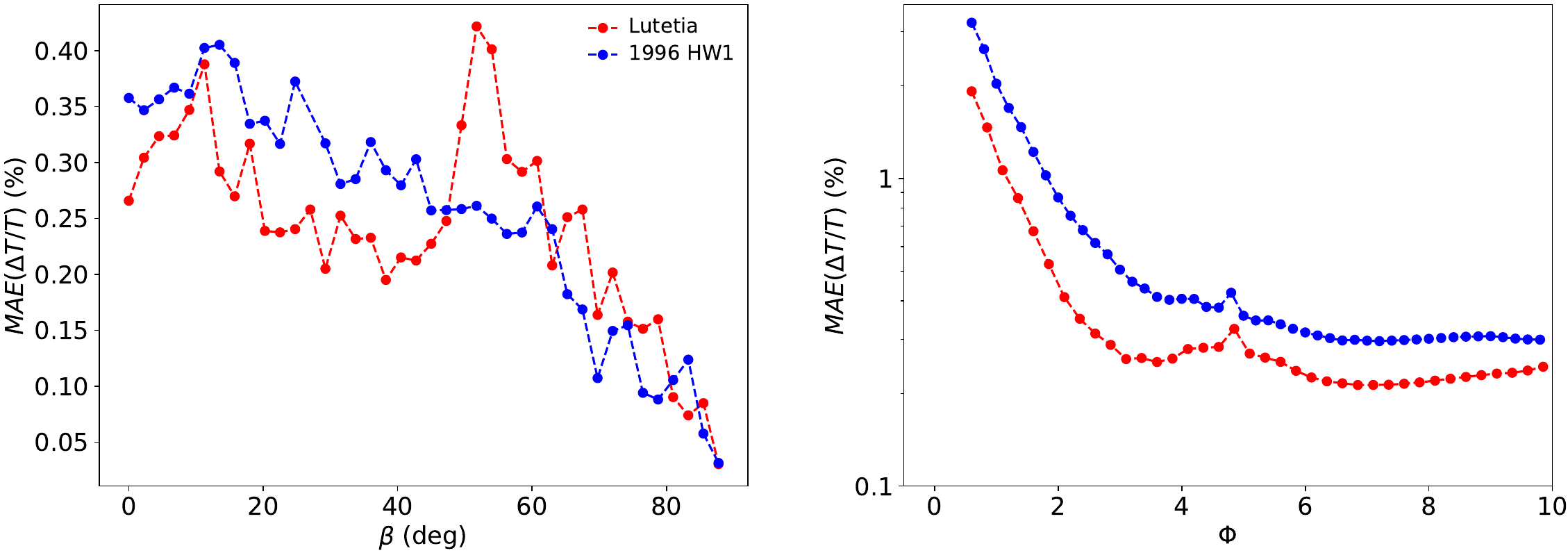}
\caption{Similar to Fig. \ref{fig:error} but with realistic shapes of irregular asteroids Lutetia and 1996 HW1. MAE represents the mean absolute error.}
\label{fig:error1}
\end{figure*}

The relative errors remain at a low level less than 1$\%$ with the majority of facets. Only a small number of facets have errors around 2$\%$. The projected shadows do have an impact on the computation accuracy, as the errors are slightly larger in the areas with significant self-shadowing effect, but still within an acceptable level. Likewise as the previous analysis, we also evaluate the MAE of $\Delta T/T$ as the functions of $\beta$ and $\Phi$, which are shown in Fig. \ref{fig:error1}. The errors are consistent with those shown in Fig. \ref{fig:error}, hence showing the strong generalization ability of DeepONet, supporting reliable calculations even with high-complexity input functions.

\section{Applications}\label{sec5:app}

As mentioned before, the short inference time by using DeepONet makes it possible for solving problems that require a large number of repeated computation of temperature distributions. In this section, we apply DeepONet to two such problems: analysis of Yarkovsky force in multi-parameter space and the orbital evolution of asteroids under precise Yarkovsky effect.

\subsection{Parameter analysis}\label{subsec5:pa}

The temperature distribution and the resulting Yarkovsky force of asteroids are closely related to their physical properties. It is crucial to understand their relationship, which can be applied to exact prediction about how the Yarkovsky force changes with a certain physical parameter \citep{Bottke2006}. In addition, it is beneficial for the problem of parameter inversion of asteroids.

Traditionally, the inversion of an asteroid’s thermal-related properties is performed by minimizing a fitting factor measuring the discrepancy between the radiation flux from simulated temperature distribution and mid-infrared observation data \citep{Müller2011,Emery2014,Yu2015}. Alternatively, when the drift of the orbital semi-major axis is measured, fitting the Yarkovsky effect with variable parameters also lead to the estimation properties \citep{Chesley2003,Vokrouhlick2008,Chesley2014,Hanus2018}.

Both methods require the exploration of a large parameter space that is usually computationally expensive. Here, we take advantage of DeepONet to make parameter analysis about Yarkovsky acceleration for a spherical asteroid with 2112 facets. We explore the distributions of transverse and radial Yarkovsky accelerations in the ($\rho$, $C$), ($\beta$, $r$) and ($R$, $\kappa$) spaces with high resolution ratio (100 $\times$ 100). In each space, the remaining parameters are fixed as default values, as shown in Table \ref{tab:paraspace}. It takes a short time to produce distributions of Yarkovsky acceleration in the three parameter spaces. In Sect \ref{subsec4:tce}, it is mentioned that using matrix operations instead of iterative methods to compute global temperature distributions under various conditions can greatly reduce time costs. The specific implementation details can be found in the code we have uploaded to GitHub\footnote{https://github.com/zsjnb7/DeepONet-for-asteroids-temperature}. In this case, the computation time for Yarkovsky acceleration in 2-dimension space can be reduced to 8 seconds.

\begin{table}[]
\footnotesize
\centering
\caption{Default setting of parameters in parameter space analysis. In each space, only the corresponding parameters change, the remaining parameters are selected from this table.}
\setlength{\tabcolsep}{4 mm}{
\begin{tabular*}{0.5\textwidth}{@{\extracolsep{\fill}}lc@{\extracolsep{\fill}}}
    \hline
        {Bond albedo $A_B$} & 0.122 \\ \hline
        {Emissivity $\epsilon$} & 0.9 \\ \hline
        {Specific heat capacity $C$ $\mathrm{(J\;K^{-1}kg^{-1})}$} & 680 \\ \hline
        {Surface density $\rho$ $\mathrm{(kg/m^3)}$} & 1500 \\ \hline
        {Bulk density $\rho_b$ $\mathrm{(kg/m^3)}$} & 1500 \\ \hline
        {Thermal conductivity $\kappa$ $\mathrm{(W\;m^{-1}K^{-1})}$} & 0.15 \\ \hline
        {Diameter $d$ $\mathrm{(km)}$} & 5.1 \\ \hline
        {Obliquity of spin axis $\beta$ $\mathrm{(deg)}$} & 0 \\ \hline
        {Heliocentric distance $r$ $\mathrm{(au)}$} & 0.6 \\ \hline
        {Rotation period $P$ $\mathrm{(h)}$} & 3.6 \\ \hline
\end{tabular*}}
\label{tab:paraspace}
\end{table} 

\begin{figure*}
\centering
\includegraphics[width=0.8\linewidth]{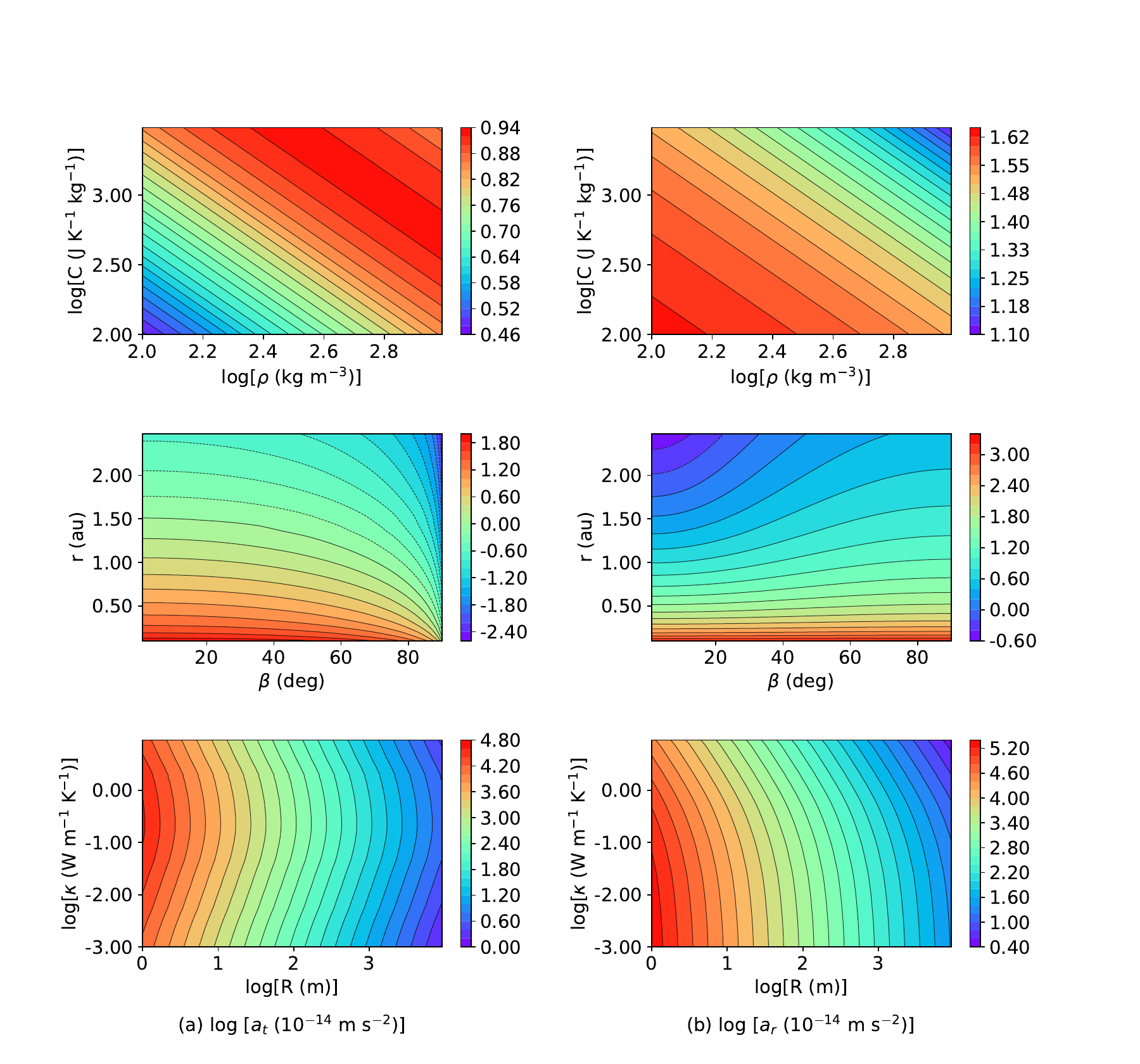}
\caption{Magnitude of (a) transverse, and (b) radial Yarkovsky accelerations produced by DeepONet for a sphere asteroid show in different parameter space, spannned by the specific heat capacity $C$, the density $\rho$, the distance from the Sun $r$, the rotational inclination $\beta$, the radius $R$ and the thermal conductivity $\kappa$. For the three situations, excepting for the two studied parameters, other parameters are fixed. The magnitude of acceleration is shown in color bar.} 
\label{fig:parameter}
\end{figure*}

Results showing in Fig. \ref{fig:parameter} indicate: (a) The transverse Yarkovsky acceleration increases with surface density and specific heat capacity, whereas the radial acceleration shows the opposite trend. (b) The Yarkovsky acceleration decreases with increasing heliocentric distance. While the transverse component decreases with increasing obliquity, the radial part shows the opposite trend. (c) Both accelerations decrease with the size of asteroid. The transverse component increases first and then decreases with thermal conductivity, and the radial component consistently decreases. We note that the dependency also reflects the influence of $\Phi$ on the transverse and radial accelerations, which can be understood from the theoretical relationships given in \citet{Vokrouhlicky2017}
\begin{equation*}
|a_t| \sim \frac{\frac{1}{2}\Phi}{1+\Phi+\frac{1}{2}\Phi^2},\;|a_r| \sim \frac{1+\frac{1}{2}\Phi}{1+\Phi+\frac{1}{2}\Phi^2}
\end{equation*}
which provide first guess for the non-linear case. These behaviours are also compatible to those found in previous works by \citet{Bottke2006} and \citet{Xu2022}. 

\subsection{Orbital evolutions under the Yarkovsky effect}\label{Sunsec5:se}

Using the traditional numerical simulations, it is difficult to consider long-term evolution of asteroids under the N-body problem with precise Yarkovsky effect. Thus, the widely used approach for treating the Yarkovsky effect in the process of discussing orbital evolution is to make linear approximation (the so-called analytical method) \citep{Vokrouhlicky19981,Vokrouhlicky1998,Vokrouhlicky1999,Bottke2006}. If irregular asteroids are considered, these errors introduced by linear approximation may be amplified \citep{Xu2022}, resulting in significant deviations. DeepONet enables fast computation of the Yarkovsky force while maintaining a certain level of accuracy, making it be possible to provide real-time precise Yarkovsky force during the orbital evolution in a N-body system.

\begin{figure}
\centering
\includegraphics[width=0.48\textwidth]{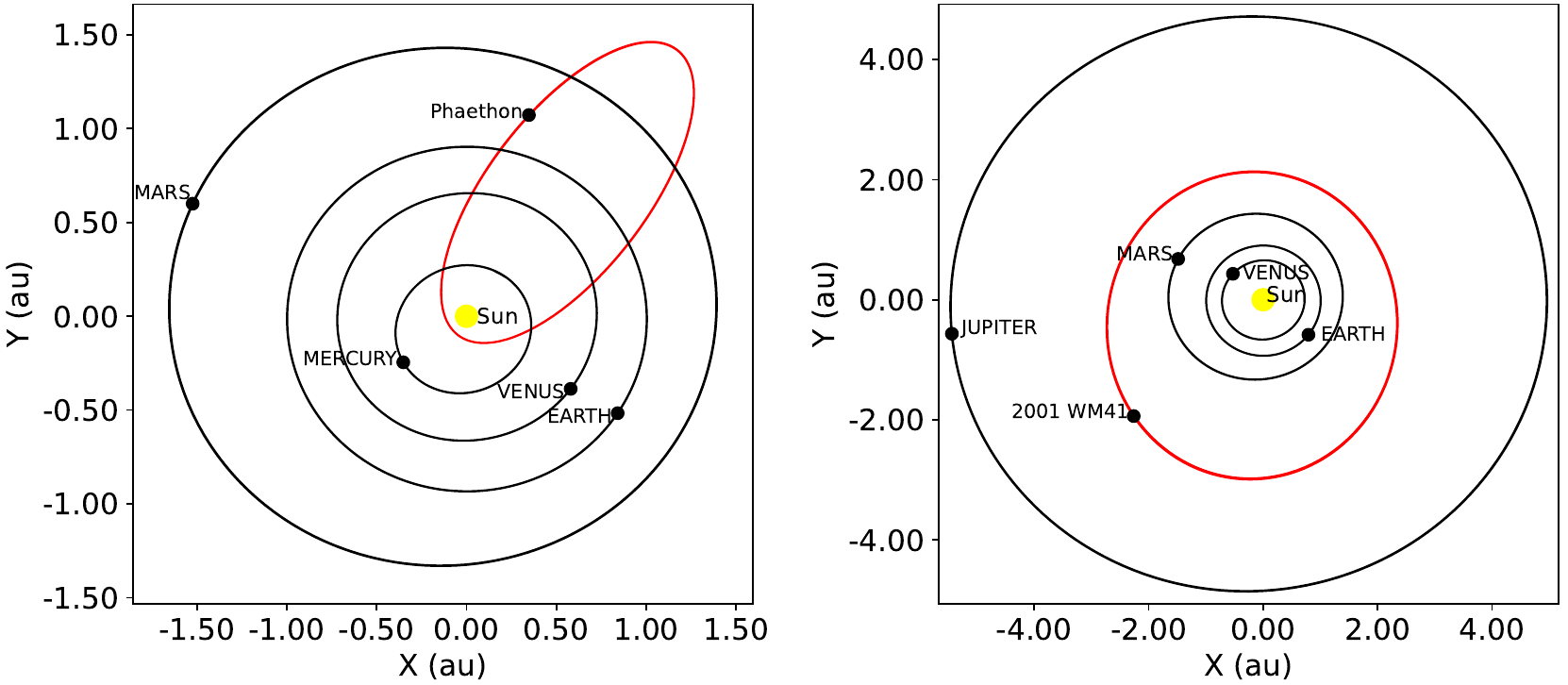}
\caption{Orbits of Phaethon (left) and 2001 WM41 (right) in the J2000.0 Earth equatorial coordinate system. The yellow points represent the Sun, the black lines are the planet orbits and the red lines represent the orbits of asteroids.} 
\label{fig:orbit}
\end{figure}

We take the near-Earth asteroid (3200) Phaethon and the main-belt asteroid (89433) 2001 WM41 as representatives. Their orbital configurations are shown in Fig \ref{fig:orbit}. As the rotation period of the two asteroids is much shorter than their orbital periods, we assumed that the surface temperature distribution reaches a dynamic equilibrium state at any point of their orbits. Therefore, it is possible to obtain instantaneous Yarkovsky force at each position along the orbit, which is applied to the orbit propagation with a N-body dynamical model including the gravitational perturbations from eight planets and the moon. The results are compared to those obtained from analytical methods (linear models mentioned in Sect. \ref{sec1:intro}). 

\begin{figure}
\centering
\includegraphics[width=0.45\textwidth]{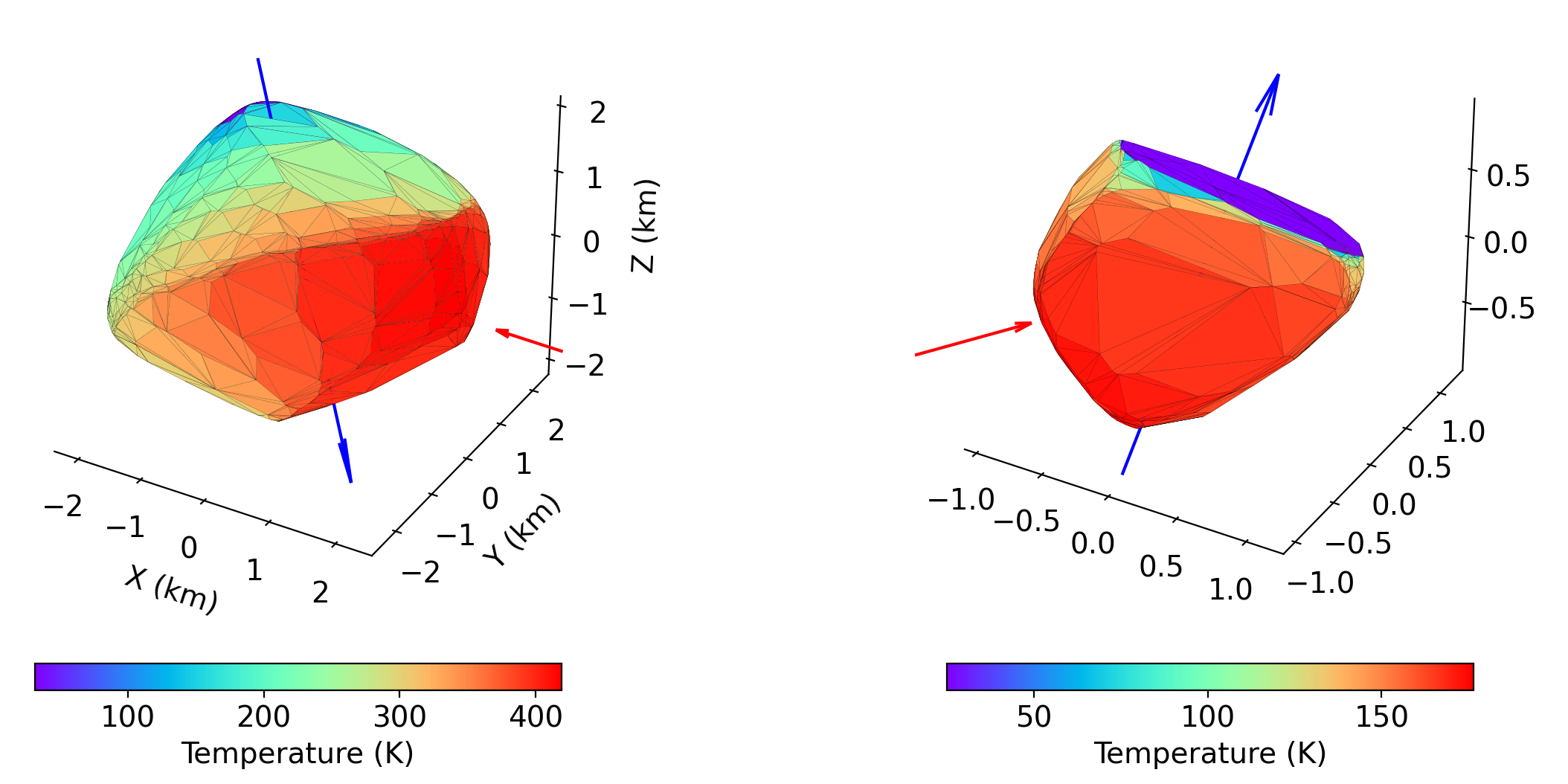}
\caption{Convex polyhedron shape models and surface temperatures of Phaethon and 2001 WM41 at the initial time. The left panel corresponds to Phaethon and the right panel is for 2001 WM41. The blue arrows represent the direction of rotation axis and the red arrows represent the direction of the Sunlight at the initial time. The surface temperatures are shown in color bar.} 
\label{fig:example}
\end{figure}

The shape models and physical parameters of Phaethon and 2001 WM41 are from DAMIT\footnote{https://astro.troja.mff.cuni.cz/projects/damit/} \citep{Hanus2016Phaethon,Hanus2016}. Their polyhedron models are shown in Fig. \ref{fig:example} and parameters are provided in Table \ref{tab:dpa}. It should be mentioned that some parameters for 2001 WM41 are unknown, and they are taken as hypothetical values. The surface density and bulk density are assumed the same. The initial time and evolution time is 2023.9.13.0 and 20,000 orbital period for Phaethon, 2024.3.31.0 and 10,000 orbital period for 2001 WM41. The initial orbital data comes from the Minor Planet Center\footnote{https://www.minorplanetcenter.net/}. 

\begin{table}[]
\footnotesize
\centering
\caption{Thermal and dynamical parameters of Phaethon and 2001 WM41.}
\setlength{\tabcolsep}{4 mm}{
\begin{tabular*}{0.5\textwidth}{@{\extracolsep{\fill}}lcc@{\extracolsep{\fill}}}
    \hline
        {Property} & {Phaethon} & {2001 WM41} \\ \hline
        {Bond albedo $A_B$} & 0.122 & 0.332 \\ \hline
        {Emissivity $\epsilon$} & 0.9 & 0.9 \\ \hline
        {Thermal inertia $\Gamma$ $\mathrm{(J\;m^{-2}s^{-1/2}K^{-1})}$} & 300 & 300 \\ \hline
        {Density $\rho$ $\mathrm{(kg/m^3)}$} & 1500 & 1500 \\ \hline
        {Diameter $d$ $\mathrm{(km)}$} & 5.1 & 2.5 \\ \hline
        {Direction of Spin axis $(\lambda,\beta)$} & $(318^{\circ},-47^{\circ})$ & $(72^{\circ},61^{\circ})$ \\ \hline
        {Rotation period $P$ $\mathrm{(h)}$} & 3.6 & 7.7 \\ \hline
\end{tabular*}}
\label{tab:dpa}
\end{table}

Because the Mercury is considered in the N-body dynamical model, the time step of numerical integration becomes very small, resulting in a staggering number of computation times for the Yarkovsky forces during orbit integration. A precomputed high-precision database is produced and called during numerical propagation. The database is generated based on the position of the Sun. Each position of the Sun is represented by spherical coordinates $r$, $\theta$, $\phi$ in the asteroid body-fixed coordinate system, where the origin is located at the center of the asteroid, z-axis is parallel to the spin axis, x-axis points to the Sun when the asteroid is on the ascending node between its orbit and equator. The database with 128$\times$128$\times$128 grids is produced in about 8 minutes with the matrix manipulation (the code we have uploaded to GitHub is designed to compute this database). We have verified the accuracy of the database by comparing the short-time orbits propagated with real-time Yarkovsky accelerations calculation and with the database.

\begin{figure*}
\centering
\includegraphics[width=0.8\linewidth]{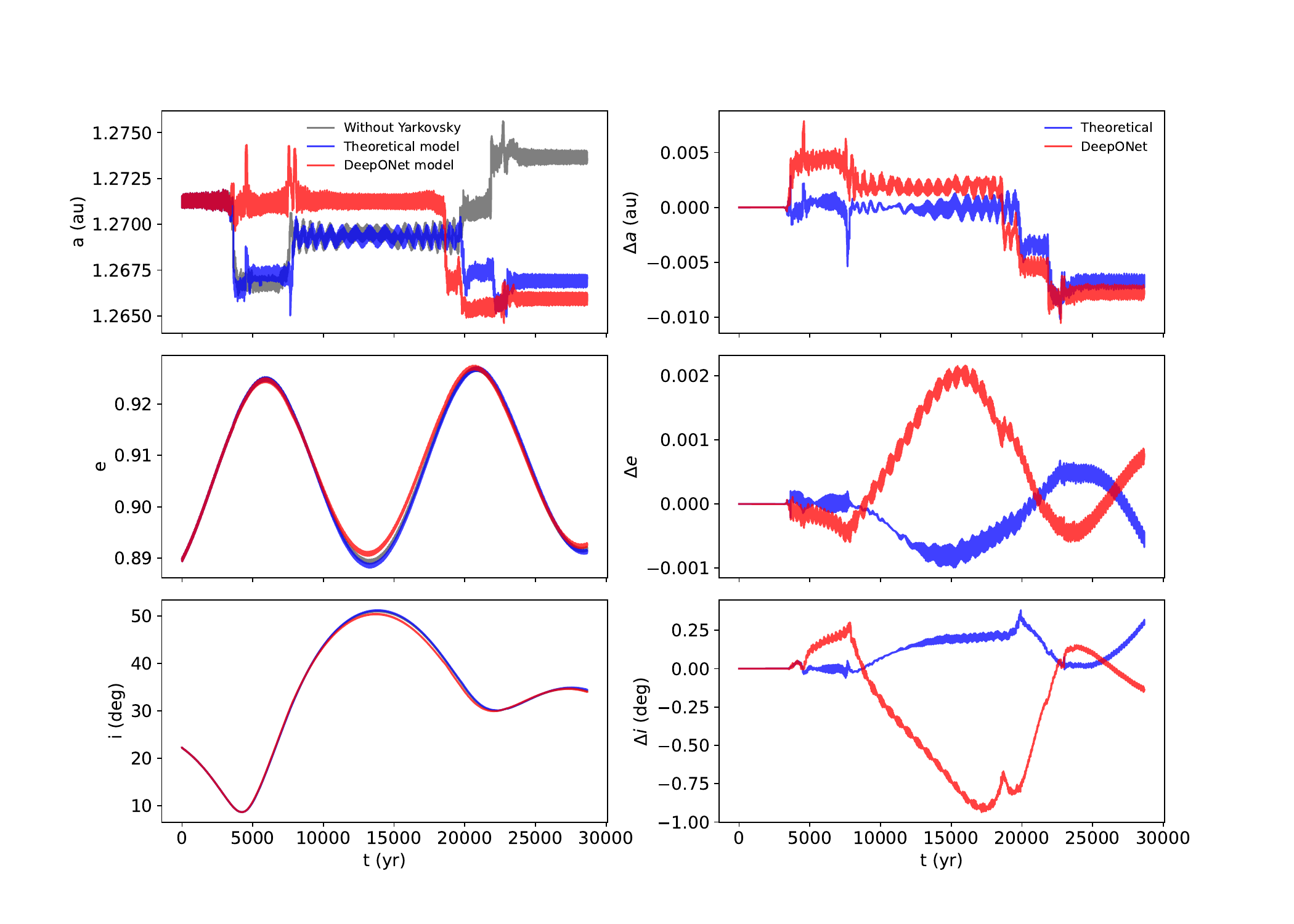}
\caption{Orbital evolution of Phaethon under different dynamical models (left-colum panels) and their differences (right panels). The evolutions of semimajor axis, eccentricity and inclination are presented in the left-column panels, where the gray lines stand for the evolution without Yarkovsky force, the blue lines represent the evolution with linear approximation of Yarkovsky force and the red lines are for the evolution with Yarkovsky force produced from DeepONet.} 
\label{fig:element_evolution1}
\end{figure*}

\begin{figure*}
\centering
\includegraphics[width=0.8\linewidth]{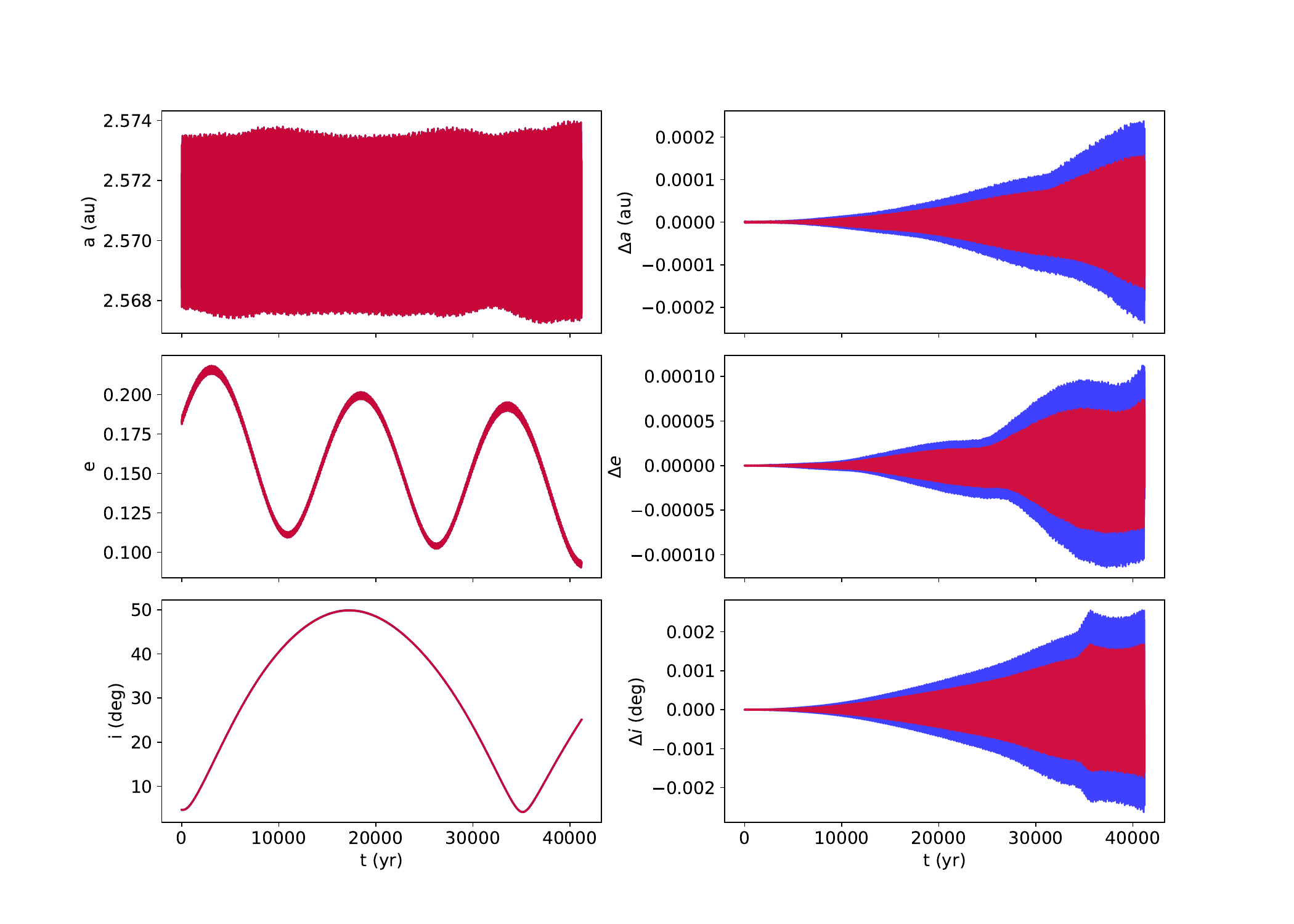}
\caption{Similar to Fig. \ref{fig:element_evolution1} but for the orbital evolution of 2001 WM41. The gray line is barely visible due to the fact that it is almost overlapping with the other two lines.}
\label{fig:element_evolution2}
\end{figure*}

Fig. \ref{fig:element_evolution1} and \ref{fig:element_evolution2} show the orbital evolutions with Yarkovsky effect developed by linear model and DeepONet. As for Phaethon, the evolution of semi-major axis is significantly influenced by Yarkovsky effect, showing visibly different behaviours under different models, even diverging in the direction. Without Yarkovsky effect, the semi-major axis of Phaethon will be larger after 30,000 years compared to its current value (showing outward migration). Both models considering Yarkovsky effect show that the semi-major axis will be smaller after 30,000 years (showing inward migration). However, evolutionary paths under the real-time Yarkovsky force computed by DeepONet and the approximate Yarkovsky force from the analytical theory is distinctly different. Regarding the inclination and eccentricity, the evolutionary trends of the three models are qualitatively similar, but also exhibit certain degrees of deviation. For 2001 WM41, the influence of Yarkovsky force is relatively small, mainly causing the amplitude of the orbital element oscillation to gradually increase but the evolution calculated by the three models is nearly identical over 40,000 years. The influence of Yarkovsky force from DeepONet is evidently smaller than that from the analytical model.

The different levels of influences of Yarkovsky effect upon orbital evolution of Phaethon and 2001 WM41 is related to the characteristics of their orbits. Phaethon is located on a highly eccentric orbit, with a small perihelion distance of 0.14 au. This causes a relatively small $\Phi$ and large $T_e$ at perihelion, resulting in large surface temperature variation with high temperature at the subsolar point of Phaethon exceeding 1000 $\mathrm{K}$. This translates to a large Yarkovsky force that could cause a significant impact on the orbital evolution. On the other hand, the close encounters to Phaethon with Venus and Earth cause highly nonlinear dynamical environment that is sensitive to initial conditions. This has been demonstrated by \citet{Hanus2016Phaethon}, where each trajectory with a different initial value corresponds to a clone variant. Therefore, the significant differences in the orbital evolution of Phaethon with and without Yarkovsky effect fundamentally arise from the perturbation (Yarkovsky effect) present in the chaotic dynamical system. Such a relatively significant Yarkovsky effect serves as a “switch” that changes the orbital evolution from one clone to another. Regarding 2001 WM41, it is located in the main belt and the Yarkovsky force is relatively small. Within the time scale of simulation, it does not play a significant role and is only coupled with N-body perturbations, leading to a slight increase of oscillating amplitude of orbital elements.

For asteroids like Phaethon in a chaotic dynamical system, there may be significant discrepancies between the actual orbital evolution and the prediction using analytical Yarkovsky models. This is due to the fact that the analytical models are based on the assumption of small temperature variation across the surface and the approximation obtained by Taylor expansion from the nonlinear terms. When the asteroid is very close to the Sun or has specific physical properties, these approximations could lead to significant temperature deviations. Furthermore, if we take into account a chaotic system or an asteroid with irregular shape, the inaccuracy in the assumption will be further amplified. However, for main-belt asteroids or others located far away from planets or sun, the differences are relatively small.

\section{Conclusion}\label{sec6:con}

In this work, we explored the possibility of using neural networks to model the surface temperature of asteroids and consider orbit evolutions of asteroids under the precise Yarkovsky effect.

We have adapted the deep operator neural network (DeepONet) by adding branch networks to learn the parameters that influence the heat conduction equation. Moreover, we have embedded the attention mechanism in the network to achieve better feature extraction. This network can learn the implicit nonlinear operators in the 1-dimensional heat conduction equation, thus achieving the goal of solving the equation. With a trained network, we can directly infer the global surface temperature distribution by inputting all facets of an asteroid as a batch.

Simulation results indicate a significant improvement in the computational speed compared to direct numerical simulation. In particular, for an asteroid with 20,000 facets, DeepONet with GPU is about $1.6 \times 10^5$ times faster than the traditional numerical method and the increase of computational cost is less steep. We assessed the accuracy of DeepONet solution by examing errors in temperature with changing obliquity and the thermal parameter. The errors can be controlled below about 2\%, averaging at about 0.2\%. Additionally, two irregular asteroids were taken to test errors in the presence of self-shadowing effect that can complicate the radiation flux function. Results show that the errors in temperature and Yarkovsky force remained at low levels.

The network with high computational efficiency and low-level error are useful for computationally heavy problems, and the typical case being the Yarkovsky effect. We tested our model on both parameter space analysis and orbital evolution related to Yarkovsky effect. Combining network with matrix computations allows for the rapid calculation of the Yarkovsky force in parameter space, which could largely facilitate any study concerning parameter inversion. 

The DeepONet model is also successfully applied in analysis of orbit evolution. We studied Phaethon and 2001 WM41 in N-body systems under Yarkovsky effects, the former one is located in a highly eccentric orbit near Earth, and the latter one is in the main belt. Simulation results indicate that the Yarkovsky effect has a significant impact on Phaethon, showing an overall decreasing trend in its semi-major axis, while the effect on 2001 WM41 is relatively weak. Furthermore, we compared the relatively accurate evolution of orbital elements with the results calculated using analytical models, suggesting that the numerical models are the best choice to solve the problems about irregular asteroids, especially Phaethon-like asteroids with large variations of surface temperature or in a chaotic dynamic system, the adoption of neural network like DeepONet makes the inclusion of precise Yarkovsky effect possible.

We anticipate a broad range of applications for this DeepONet-based 
thermophysical modeling in analysing problems, such as multidimensional parameter inversion, long-term orbital evolution under the combination of instantaneous Yarkovsky and YORP effects, etc. Similar AI-based approach can be extended to more complex physical processes, such as the thermophysical modeling of comets.

\begin{acknowledgements}
\textit {The authors thank Dr. Vokrouhlicky for insightful review comments that helped significantly improved the manuscript. SJZ wishes to thank Drs. Yining Zhang and Yangbo Xu for helpful discussions about the thermophysical model of asteroids. XS thanks members of the ISSI International Team “Timing and Processes of Planetesimal Formation and Evolution” for inspirational discussions. This work is financially supported by the National Natural Science Foundation of China (Nos. 12073011 and 12233003) and the National Key R\&D Program of China (No. 2019YFA0706601).}
\end{acknowledgements}

\bibliographystyle{aa}
\bibliography{references}
\end{document}